\documentclass[aps,prl,twocolumn,showpacs,superscriptaddress]{revtex4-1}

\usepackage{amssymb}
\usepackage{natbib}
\usepackage{amsmath}    % need for subequations
\usepackage{graphicx}   % need for figures
\usepackage{verbatim}   % useful for program listings
\usepackage{color}      % use if color is used in text
\usepackage{subfigure}  % use for side-by-side figures
\usepackage{hyperref}   % use for hypertext links, including those to
\usepackage{epstopdf}

\begin{document}

\title{The Spin-Polarization of the Split Kondo State}
\author{Kirsten~von~Bergmann}
\affiliation{Department of Physics, University of Hamburg, 20355 Hamburg, Germany}
\author{Markus~Ternes}
\email{m.ternes@fkf.mpg.de}
\affiliation{Max-Planck Institute for Solid State Research, 70569 Stuttgart, Germany}
\author{Sebastian~Loth}
\affiliation{Max-Planck Institute for the Structure and Dynamics of Matter, 22761 Hamburg, Germany}
\affiliation{Max-Planck Institute for Solid State Research, 70569 Stuttgart, Germany}
\author{Christopher~P.~Lutz}
\affiliation{IBM Research Division, Almaden Research Center, San Jose, California 95120, USA}
\author{Andreas~J.~Heinrich}
\affiliation{IBM Research Division, Almaden Research Center, San Jose, California 95120, USA}

\date{\today}

\begin{abstract}
Spin-resolved scanning tunneling microscopy is employed to quantitatively determine the spin-polarization of the magnetic field-split Kondo state. Tunneling conductance spectra of a Kondo-screened magnetic atom are evaluated within a simple model taking into account inelastic tunneling due to spin excitations and two Kondo peaks positioned symmetrically around the Fermi energy. We fit the spin state of the Kondo-screened atom with a Spin Hamiltonian independent of the Kondo effect and account for Zeeman-splitting of the Kondo peak in the magnetic field. We find that the width and the height of the Kondo peaks scales with the Zeeman energy. Our observations are consistent with full spin-polarization of the Kondo peaks, i.e., a majority spin peak below the Fermi energy and a minority spin peak above.
\end{abstract}

\pacs{72.15.Qm, 68.37.Ef}

\maketitle

The Kondo effect was discovered experimentally nearly one century ago and starting from the 1960s theory has been employed to unravel its origin and properties~\cite{Hewson1997}. It arises from the screening of a localized magnetic moment by host electrons, which leads to a Fermi-level resonance in the density of states. Kondo physics in the absence of a magnetic field has been studied extensively~\cite{Pustilnik2004,Ternes2009}, whereas the precise behavior of the Kondo state in a magnetic field is less studied~\cite{Costi2000,Otte2008,Quay2007}. The coupling of the localized spin to the environment sets the relevant energy scale which is typically referred to as Kondo temperature~$T_{\mathrm K}$. For magnetic fields exceeding this energy the Kondo resonance splits. However, the spin-resolved properties of this split Kondo state and in particular the amount of spin-polarization of the two resulting peaks remains elusive~\cite{Patton2007,Seridonio2009}. While there is one spin-resolved measurement of a split Kondo-state~\cite{Fu2012}, the asymmetry of the peaks was not studied systematically and a comprehensive picture is missing.

The Kondo effect of a single atom or molecule on a surface can be probed with scanning tunneling spectroscopy~(STS)~\cite{Ternes2009}. When the magnetic atom exhibiting Kondo scattering is only weakly coupled to the conduction electrons of the substrate a perturbative description can be used~\cite{Appelbaum1967} and a logarithmic peak at the Fermi energy is detected~\cite{Zhang2013}. For stronger hybridization with the substrate the arising correlations lead to a change in the density of states~\cite{Costi2000}, which is typically described by a Lorentzian or a Frota function~\cite{Frota.PhysRevB.45.1096}. An additional linear tunnel channel gives rise to interference between different paths, leading to an asymmetric, Fano-like, lineshape~\cite{Madhavan1998,Prueser2011}. Systems with a spin $S > 1/2$ can show the Kondo effect when the magnetocrystalline anisotropy leads to a degenerate $m = \pm 1/2$ ground state~\cite{Otte2008}. Then the resonance at zero bias is accompanied by steps in the tunnel spectra due to inelastic electron tunneling that excites the spin at finite energy. Inelastic tunnel spectroscopy has been employed to investigate such spin excitations in single atoms, molecules, and small clusters on ultra-thin insulating layers, semiconductors, and metals~\cite{Heinrich2004,Hirjibehedin2006,Otte2008,Khajetoorians.N2010,Khajetoorians.PRL2011,Kahle2012}. All these experiments were interpreted within a model spin Hamiltonian taking into account the spin, the $g$-factor, and the magnetocrystalline anisotropy, reproducing the magnetic field dependence of the observed steps in the tunnel spectra.

In this letter we quantitatively determine the spin-polarization of the magnetic field-split Kondo state with spin-resolved STS in an external magnetic field. As a Kondo system we use a Co $S = 3/2$ atom on a thin insulating Cu$_2$N layer, decoupling it from a metallic Cu(001) surface. The arising crystal field splitting leads to a $m = \pm 1/2$ ground state of the adatom~\cite{Otte2008}. We fit our experimental STS data using a model that treats the contributions to the differential conductance from inelastic spin flip excitations and Kondo effect separately. Via the spin flip excitations from the $m = 1/2$ to $3/2$ states we can determine the magnetic properties of the adatom independent from the Kondo resonance. Using the experimentally determined spin-polarization of the tip we can directly deduce that the split Kondo state is fully polarized.

\begin{figure*}[]
\includegraphics[width=2\columnwidth]{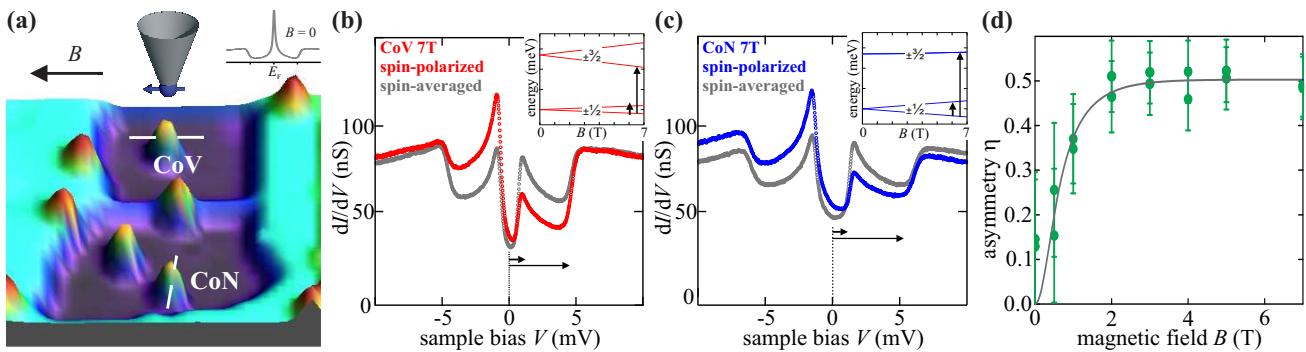}
\caption{(color online). \textbf{(a)}~Setup of the spin-resolved STM experiment on individual Co atoms on Cu$_2$N/Cu(001) in an in-plane magnetic field, STM topography image (10\,nm $\times$ 10\,nm, $V=+10$\,mV, $I=0.1$\,nA); the right inset shows a d$I$/d$V$~spectrum of a Co atom at $B=0$. \textbf{(b)},\textbf{(c)}~Spin-averaged (gray) and spin-resolved (color) tunnel spectra of Co atoms with magnetic field of 7\,T applied along (CoV) and perpendicular (CoN) to the hard anisotropy axis, respectively (stabilization setpoint $\sigma = 80$\,nS, $V=+25$\,mV, $I=2$\,nA). Insets sketch the two relevant spin flip excitations as a function of magnetic field. \textbf{(d)}~The tip's spin polarization, deduced from measurements of a Mn atom on Cu$_2$N, as a function of applied magnetic field; the solid line is a fit accounting for the paramagnetic behavior of the Mn atoms on tip and sample by Brillouin functions for spins with $S = 5/2, g = 2, T = 0.5\,$K, $\eta_{\rm sample} = 1$, $\eta_{\rm tip} ^{\rm max} = 0.5$.}
\label{fig1}
\end{figure*}

Figure 1(a) sketches the experimental setup of our spin-polarized scanning tunneling microscopy (SP-STM) measurements~\cite{Wiesendanger2009,Loth.NP2010} of individual Co atoms on Cu$_2$N/Cu(001)~\cite{Otte2008}. Co atoms adsorb on the Cu binding site of the Cu$_2$N and have been found to have a spin of $S = 3/2$ with a hard anisotropy axis in the surface plane, perpendicular to the bonds to neighboring N atoms. Two chemically equivalent Cu binding sites exist with the uniaxial hard anisotropy axes perpendicular to each other (white lines in Fig.\,1(a)). A magnetic field $B$ applied in the surface plane gives rise to two inequivalent Co atoms on the surface regarding the direction of the magnetic field with respect to the Co magnetic quantization axis: magnetic field parallel to the hard anisotropy axis (denoted CoV) and perpendicular to it (denoted CoN). This enables measurements of two Co atoms in inequivalent magnetic field directions with the same spin-polarized tip. Tunnel spectra were acquired at the STM temperature of 0.5\,K by recording the differential conductance as a function of sample bias, d$I$/d$V$($V$), with the tip positioned above isolated atoms. The  tunnel spectra in the absence of a magnetic field exhibit two distinct features; see inset in Fig.\,1(a): a Kondo resonance at $E_{\rm F}$ and spin flip excitations appearing as steps at about $\pm 5$\,mV. In magnetic field the Kondo resonance splits and a double-peak emerges (Fig.\,1(b),(c)). In addition the energy of the spin excitation shifts and another low energy spin excitation due to a transition between $m= \pm 1/2$ states becomes possible, which is superimposed on the Kondo peaks. Magnetic fields applied along different crystallographic directions result in different amounts of splitting due to the magnetic anisotropy of the Co atoms~\cite{Otte2008}, compare insets in Fig.\,1(b) and (c) for CoV and CoN. When measured with a non--spin-polarized (spin-averaging) tip the spectra are mirror-symmetric with respect to 0\,V sample bias, whereas a spin-polarized tip leads to different step and peak heights for positive and negative bias, compare gray and colored spectra in Fig.\,1(b),(c). The asymmetry of the step heights for the spin excitations is due to the spin polarization of the tunnel current in conjunction with the selections rules of the spin flip transition~\cite{Loth.NP2010}. The role of the Kondo effect for the asymmetry of the spectra is unclear and at the heart of this investigation.

The spin-polarized tip was prepared by picking up Mn atoms from the surface with the STM tip~\cite{Loth.NP2010}. Since the measured spin-polarization in a tunnel experiment is the product of sample and tip spin polarization, $\eta^{\rm eff} = \eta_{\rm sample} \cdot \eta_{\rm tip}$, it is crucial to characterize the degree of spin-polarization of the tip. This was done by spin-resolved inelastic tunnel spectroscopy of individual Mn atoms on the same surface. Mn atoms on Cu$_2$N show one spin flip excitation at about 1\,mV~\cite{Hirjibehedin2006}. When measured with a spin-polarized tip the heights of the inelastic steps at positive and negative voltage ($h^+$ and $h^-$) differ~\cite{Loth.NP2010}. The asymmetry of the step heights $\eta_{\rm step}^{\rm eff}  = (h^- - h^+)/(h^- + h^+)$ for two individual Mn atoms is measured as a function of magnetic field. Due to the small magnetocrystalline anisotropy of Mn the nominal spin-polarization of the step is $\eta_{\rm Mn} = 1$. In this case the experimental $\eta^{\rm eff}$ is a quantitative measure of the tip spin polarization. We find that the magnetic-field dependence of the tip's spin polarization is consistent with paramagnetic behavior of the Mn atoms on the sample and the metallic tip. Hence magnetic field-dependent spin polarization of tip and sample are well described by Brillouin functions (solid line in Fig.\,1(d)). In the following we use this functional dependence to account for the field dependence of the tip's spin polarization.

\begin{figure}[]
\includegraphics[width=\columnwidth]{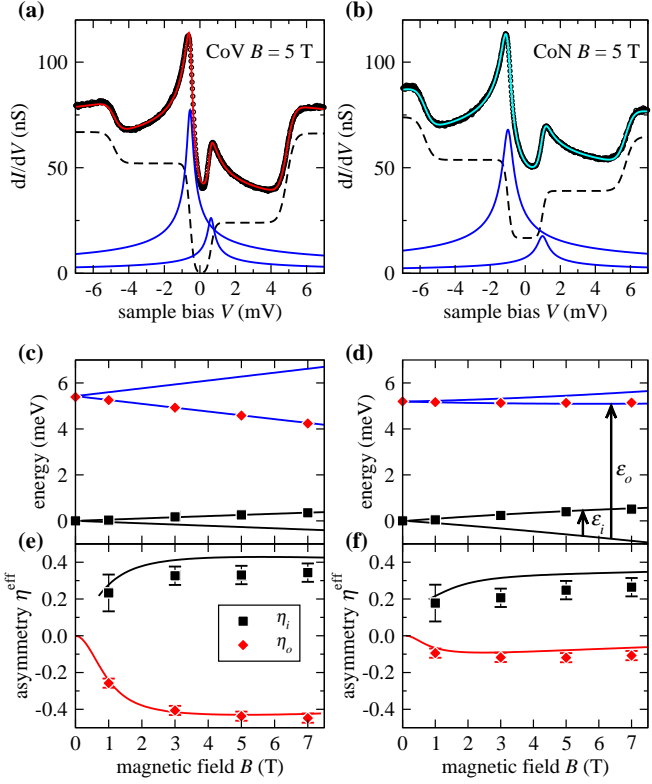}
\caption{(color online). Decomposition of the d$I$/d$V$~spectra into spin flip excitation and Kondo related contributions. \textbf{(a)},\textbf{(b)}~Experimental data (circles) measured on a CoV and a CoN atom at $B=5$\,T (stabilization setpoint $\sigma = 80$\,nS, $V=+25$\,mV, $I=2$\,nA). Upper solid line: best fit using the sum of an asymmetric double step function (dashed line) and two Frota functions (lower solid lines), according to Eq.\,(1) of the main text. \textbf{(c)},\textbf{(d)}~Energies $\varepsilon_{i,o}$ and \textbf{(e)},\textbf{(f)}~effective asymmetries of the two steps. Symbols are fits to the data, lines are calculated with the model spin Hamiltonian using $D=2.6 \pm 0.1~(2.5 \pm 0.1)$\,meV, $E=0.45 \pm 0.05~(0.40 \pm 0.05)$\,meV, and $g=2.1 \pm 0.1~(2.4 \pm 0.1)$ for the CoV (CoN) atom.}
\label{fig2}
\end{figure}

To determine the spin polarization of the split Kondo state quantitatively it is necessary to disentangle the spin flip excitation features and the Kondo
resonance in the tunnel spectra. We fit the experimental data with a simple model: we consider for each tunnel spectrum two spin flip transitions and model
the Kondo peaks by Frota functions; see experimental and decomposed spectra at $B = 5$\,T in Fig.\,2(a),(b) as an example. The full spectrum is therefore
\begin{eqnarray}
\sigma(eV)&=&\sigma_0+\\ \nonumber
& &h_i^-\theta(\varepsilon_i+eV)+h_i^+\theta(\varepsilon_i-eV)+\\ \nonumber
& &h_o^-\theta(\varepsilon_o+eV)+h_o^+\theta(\varepsilon_o-eV)+\\ \nonumber
& &h_K^-g(\varepsilon_K+eV,\Gamma_K)+h_K^+g(\varepsilon_K-eV,\Gamma_K),
\end{eqnarray}
where $\theta$ is the temperature-broadened step function
\begin{equation}
\theta(\varepsilon)=\frac{1+\left(x-1\right)\exp(x) } {\left(\exp(x)-1\right)^2},
\end{equation}
with $x=\frac{\varepsilon}{k_BT}$, and $g$ is the Frota function~\cite{Zitko2009,Prueser2011,Zhang2013}
\begin{equation}
g(\varepsilon,\Gamma_F)=\Re\left(\sqrt{\frac{i\Gamma_F}{i\Gamma_F+\varepsilon}} \right),
\end{equation}
where the half-width at half-maximum of the peak is given by $\Gamma_K=2.54\,\Gamma_F$~\cite{Frota.PhysRevB.45.1096}. At temperatures  $T\ll T_K$ and $B=0$ it is directly related to the Kondo temperature $T_K=\Gamma_K^0/k_B$ via the Boltzmann constant~$k_B$. The 'outer' step at around $\varepsilon_o = \pm5$\,mV with height $h_o$ corresponds to the excitation from �1/2 to �3/2 states; the 'inner' step with $h_i$ occurs between the $\pm1/2$ states and is found at an energy
$\varepsilon_i < 1\,$meV. Based on previous work~\cite{Otte2008} we introduce the constraint that the Kondo peak energy is identical to the energy of the
$\pm 1/2$ transition, $\varepsilon_K = \varepsilon_i$, and fit all remaining parameters allowing for a vertical offset $\sigma_0$.

The step energies $\varepsilon_o$ and $\varepsilon_i$ resulting from the fits are displayed as symbols within the energy level graphs for both a CoV and a CoN atom in Fig.\,2(c),(d). They now serve as input parameters for a calculation with a model spin Hamiltonian for each Co atom accounting for Zeeman splitting, uniaxial, and transverse magnetic anisotropy,
\begin{equation}
\hat{H}=g\mu_B \vec{B} \hat{\bf S}+D\hat{S}^2_z+E(\hat{S}^2_x+\hat{S}^2_y).
\label{Hamiltonian}
\end{equation}
The solid lines in Fig.\,2(c),(d) show the magnetic field dependence of $\varepsilon_o$ and $\varepsilon_i$ for the derived values of the magnetic anisotropy parameters $D$, $E$ and the $g$-factor, as given in the caption of Fig.\,2. The solid lines in Fig.\,2(e),(f) are the calculated effective step asymmetries for the CoN and CoV atoms: In our model $\eta^{\rm eff}$ is the product of the tip polarization as determined in Fig.\,1(d), and the polarization of the spin excitation calculated from the transition matrix elements for the Spin Hamiltonian~\cite{Loth.NJP.2010}. Whereas the measured step asymmetry $\eta_{o}$ extracted from the spectra (symbols) agrees well with the calculated effective asymmetry, the experimental values of $\eta_{i}$ are slightly smaller than the calculated ones. This is possibly caused by a competition between spin flip excitation and Kondo scattering. Considering the overall agreement between the fits to the tunnel spectra and the calculation, we conclude that we can capture the contributions from inelastic spin flip excitations with this simple model spin Hamiltonian, similar to previous work~\cite{Heinrich2004,Hirjibehedin2006,Otte2008,Khajetoorians.N2010,Khajetoorians.PRL2011,Kahle2012}.

\begin{figure}[]
\includegraphics[width=\columnwidth]{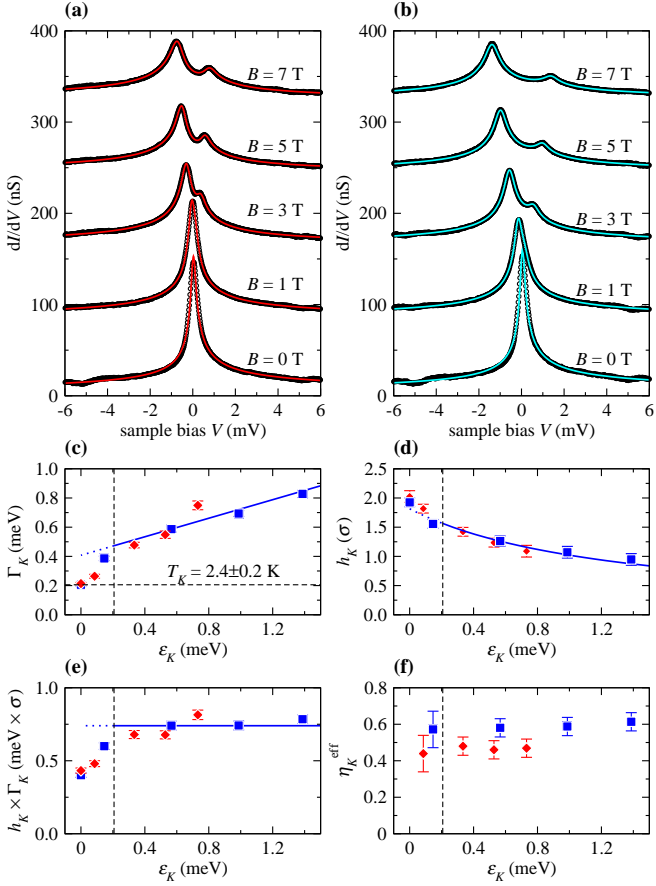}
\caption{(color online). The splitting and the polarization of the Kondo peak. \textbf{(a)},\textbf{(b)}~Kondo-related experimental data (circles) and fit with two Frota functions (solid lines) for CoV and CoN (curves in magnetic field are shifted vertically for better visualization). Fit results for \textbf{(c)}~the width $\Gamma_K$, \textbf{(d)}~the peak height $h_K$, \textbf{(e)}~the area $\Gamma_K \cdot h_K$, and \textbf{(f)}~the experimental peak asymmetry $\eta_{K}^{\rm eff}$ plotted against the splitting energy $\varepsilon_K$. The horizontal and vertical dashed lines in (c)-(f) mark the characteristic Kondo energy.}
\label{fig3}
\end{figure}

We now turn to our initial question about the spin polarization of the split Kondo peak. Figures 3(a),(b) present the Kondo-related spectra, i.e.\ the difference between the original data and the spin flip excitation contributions as fitted in Fig.\,2. It becomes apparent, that the asymmetry at $\pm \varepsilon_K$ is not only due to the polarization of the spin flip excitation step at $\varepsilon_i$, but that indeed there is a significant asymmetry of the Kondo peaks. The fitting parameters related to the Kondo feature (cf.\ Eqs.\,1 and 3) are displayed in Fig.\,3(c)-(f). We find that the peak width $\Gamma_K$ increases, Fig.\,3(c), and the peak height $h_K$ drops, Fig.\,3(d), with increasing magnetic field for both CoV and CoN. Interestingly the behavior of the Co atoms seems to be related to the splitting energy as the data for CoV and CoN fall on top of each other when plotted versus the peak energy $\varepsilon_K$. From the peak width at $B=0$ ($\Gamma_K^0$) the Kondo temperature can be extracted to $T_{\rm K} =2.4 \pm 0.2\,$K, which is equivalent to an energy of 0.21\,meV. When the splitting of the Kondo feature due to the applied magnetic field exceeds this energy (cf.\ dashed vertical lines), we observe linear behavior for the peak width, which surprisingly follows a simple equation:
\begin{equation}
\Gamma_K(\varepsilon_K)=(2\pm0.1)\Gamma_K^0+(1\pm0.03)\frac{1}{\pi} \varepsilon_K.
\end{equation}
Furthermore, we observe a $1/\varepsilon_K$ dependence for the peak height. Consequently, the area under the Kondo peak remains constant irrespective of magnetic field up to 7\,T (Fig.\,3(e)). Figure 3(f) demonstrates that the effective polarization of the Kondo peak $\eta_{\rm K}^{\rm eff}$ for both CoV and CoN stays constant and is close to 0.5.
The small systematic offset between the CoV and CoN polarization may be related to higher order effects in the inelastic tunneling~\cite{Zhang2013}.

\begin{figure}[]
\includegraphics[width=\columnwidth]{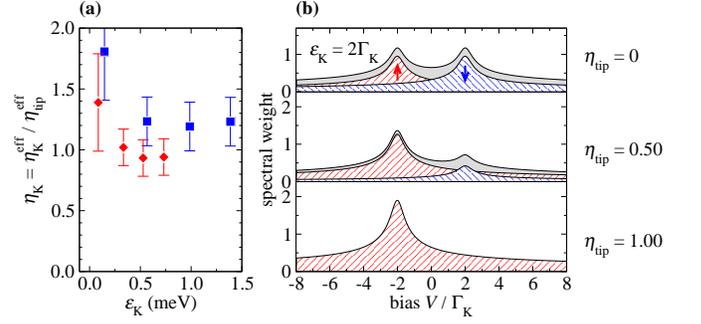}
\caption{(color online). \textbf{(a)}~The Kondo polarization $\eta_K$ is approximately one for both Co atoms independent of $\varepsilon_K$. \textbf{(b)}~Simulated spectra with $\eta_K=1$ and a magnetic field which splits the Kondo state by $\varepsilon_K=2\Gamma_K$ for different tip spin polarizations; gray is the sum of majority and minority states, i.e.\ the experimentally measured spectrum.}
\label{fig4}
\end{figure}

In Fig.\,4(a) we plot the spin-polarization of the Kondo peak $\eta_{\rm K}$ and find that it scatters around one. This key finding indicates that the Kondo feature is fully spin-polarized, meaning that the peak at negative sample bias, i.e.\ occupied density of states, is entirely in the spin-up density of states (majority spin states) and the peak at positive bias, i.e. unoccupied states, is entirely in the spin-down density of states (minority spin states). The implication for a tunnel spectroscopy experiment in an external magnetic field is illustrated in Fig.\,4(b): when the tip polarization is 0, two Kondo peaks with equal height are measured, whereas a fully spin-polarized tip with $\eta_{\rm {tip}}=1$ would lead to a single peak at $-\varepsilon_K$.

To conclude, our experiments show that the magnetic field-split Kondo state can be described by two independent fully spin-polarized peaks, one with spin-up and the other one with spin down density of states. The two peaks shift by the Zeeman energy according to the model spin Hamiltonian. For magnetic fields that exceed the characteristic Kondo energy, the width of the peaks increases linearly with the splitting energy, and together with a decreasing height this leads to a constant spectral weight of the fully spin-polarized peaks. These experimental results can serve as a reference for further theory studies of the spin-resolved behavior of the Kondo effect in an external magnetic field. In addition the field-split Kondo state could be exploited to serve as a magnetic probe in transport measurements, similar to the fully spin-polarized magnetic field split superconducting state~\cite{Meservey1994}. Experimentally, this could be realized, for example, by attaching a magnetic molecule that exhibits a Kondo resonance to the tip apex. The Kondo resonance would then act as an energy-dependent spin filter for quantitative spin-resolved STM measurements.

\begin{acknowledgments}
K.v.B.\ acknowledges financial support from the German research foundation (DFG Forschungsstipendium), S.L. from the Alexander von Humboldt foundation, AJH and CPL from the Office of Naval Research.
\end{acknowledgments}


\begin{thebibliography}{24}%
\makeatletter
\providecommand \@ifxundefined [1]{%
 \@ifx{#1\undefined}
}%
\providecommand \@ifnum [1]{%
 \ifnum #1\expandafter \@firstoftwo
 \else \expandafter \@secondoftwo
 \fi
}%
\providecommand \@ifx [1]{%
 \ifx #1\expandafter \@firstoftwo
 \else \expandafter \@secondoftwo
 \fi
}%
\providecommand \natexlab [1]{#1}%
\providecommand \enquote  [1]{``#1''}%
\providecommand \bibnamefont  [1]{#1}%
\providecommand \bibfnamefont [1]{#1}%
\providecommand \citenamefont [1]{#1}%
\providecommand \href@noop [0]{\@secondoftwo}%
\providecommand \href [0]{\begingroup \@sanitize@url \@href}%
\providecommand \@href[1]{\@@startlink{#1}\@@href}%
\providecommand \@@href[1]{\endgroup#1\@@endlink}%
\providecommand \@sanitize@url [0]{\catcode `\\12\catcode `\$12\catcode
  `\&12\catcode `\#12\catcode `\^12\catcode `\_12\catcode `\%12\relax}%
\providecommand \@@startlink[1]{}%
\providecommand \@@endlink[0]{}%
\providecommand \url  [0]{\begingroup\@sanitize@url \@url }%
\providecommand \@url [1]{\endgroup\@href {#1}{\urlprefix }}%
\providecommand \urlprefix  [0]{URL }%
\providecommand \Eprint [0]{\href }%
\providecommand \doibase [0]{http://dx.doi.org/}%
\providecommand \selectlanguage [0]{\@gobble}%
\providecommand \bibinfo  [0]{\@secondoftwo}%
\providecommand \bibfield  [0]{\@secondoftwo}%
\providecommand \translation [1]{[#1]}%
\providecommand \BibitemOpen [0]{}%
\providecommand \bibitemStop [0]{}%
\providecommand \bibitemNoStop [0]{.\EOS\space}%
\providecommand \EOS [0]{\spacefactor3000\relax}%
\providecommand \BibitemShut  [1]{\csname bibitem#1\endcsname}%
\let\auto@bib@innerbib\@empty
%</preamble>
\bibitem [{\citenamefont {Hewson}(1997)}]{Hewson1997}%
  \BibitemOpen
  \bibfield  {author} {\bibinfo {author} {\bibfnamefont {A.~C.}\ \bibnamefont
  {Hewson}},\ }\href@noop {} {\emph {\bibinfo {title} {The Kondo Problem to
  Heavy Fermions}}}\ (\bibinfo  {publisher} {Cambridge University Press},\
  \bibinfo {year} {1997})\BibitemShut {NoStop}%
\bibitem [{\citenamefont {Pustilnik}\ and\ \citenamefont
  {Glazman}(2004)}]{Pustilnik2004}%
  \BibitemOpen
  \bibfield  {author} {\bibinfo {author} {\bibfnamefont {M.}~\bibnamefont
  {Pustilnik}}\ and\ \bibinfo {author} {\bibfnamefont {L.}~\bibnamefont
  {Glazman}},\ }\href {\doibase 10.1088/0953-8984/16/16/R01} {\bibfield
  {journal} {\bibinfo  {journal} {J. Phys.: Condens. Matter}\ }\textbf
  {\bibinfo {volume} {16}},\ \bibinfo {pages} {R513} (\bibinfo {year}
  {2004})}\BibitemShut {NoStop}%
\bibitem [{\citenamefont {Ternes}\ \emph {et~al.}(2009)\citenamefont {Ternes},
  \citenamefont {Heinrich},\ and\ \citenamefont {Schneider}}]{Ternes2009}%
  \BibitemOpen
  \bibfield  {author} {\bibinfo {author} {\bibfnamefont {M.}~\bibnamefont
  {Ternes}}, \bibinfo {author} {\bibfnamefont {A.~J.}\ \bibnamefont
  {Heinrich}}, \ and\ \bibinfo {author} {\bibfnamefont {W.-D.}\ \bibnamefont
  {Schneider}},\ }\href {http://stacks.iop.org/0953-8984/21/i=5/a=053001}
  {\bibfield  {journal} {\bibinfo  {journal} {Journal of Physics: Condensed
  Matter}\ }\textbf {\bibinfo {volume} {21}},\ \bibinfo {pages} {053001}
  (\bibinfo {year} {2009})}\BibitemShut {NoStop}%
\bibitem [{\citenamefont {Costi}(2000)}]{Costi2000}%
  \BibitemOpen
  \bibfield  {author} {\bibinfo {author} {\bibfnamefont {T.~A.}\ \bibnamefont
  {Costi}},\ }\href {\doibase 10.1103/PhysRevLett.85.1504} {\bibfield
  {journal} {\bibinfo  {journal} {Phys. Rev. Lett.}\ }\textbf {\bibinfo
  {volume} {85}},\ \bibinfo {pages} {1504} (\bibinfo {year}
  {2000})}\BibitemShut {NoStop}%
\bibitem [{\citenamefont {Otte}\ \emph {et~al.}(2008)\citenamefont {Otte},
  \citenamefont {Ternes}, \citenamefont {von Bergmann}, \citenamefont {Loth},
  \citenamefont {Brune}, \citenamefont {Lutz}, \citenamefont {Hirjibehedin},\
  and\ \citenamefont {Heinrich}}]{Otte2008}%
  \BibitemOpen
  \bibfield  {author} {\bibinfo {author} {\bibfnamefont {A.~F.}\ \bibnamefont
  {Otte}}, \bibinfo {author} {\bibfnamefont {M.}~\bibnamefont {Ternes}},
  \bibinfo {author} {\bibfnamefont {K.}~\bibnamefont {von Bergmann}}, \bibinfo
  {author} {\bibfnamefont {S.}~\bibnamefont {Loth}}, \bibinfo {author}
  {\bibfnamefont {H.}~\bibnamefont {Brune}}, \bibinfo {author} {\bibfnamefont
  {C.~P.}\ \bibnamefont {Lutz}}, \bibinfo {author} {\bibfnamefont {C.~F.}\
  \bibnamefont {Hirjibehedin}}, \ and\ \bibinfo {author} {\bibfnamefont
  {A.~J.}\ \bibnamefont {Heinrich}},\ }\href {\doibase 10.1038/nphys1072}
  {\bibfield  {journal} {\bibinfo  {journal} {Nature Physics}\ }\textbf
  {\bibinfo {volume} {4}},\ \bibinfo {pages} {847} (\bibinfo {year}
  {2008})}\BibitemShut {NoStop}%
\bibitem [{\citenamefont {Quay}\ \emph {et~al.}(2007)\citenamefont {Quay},
  \citenamefont {Cumings}, \citenamefont {Gamble}, \citenamefont {Picciotto},
  \citenamefont {Kataura},\ and\ \citenamefont {Goldhaber-Gordon}}]{Quay2007}%
  \BibitemOpen
  \bibfield  {author} {\bibinfo {author} {\bibfnamefont {C.~H.~L.}\
  \bibnamefont {Quay}}, \bibinfo {author} {\bibfnamefont {J.}~\bibnamefont
  {Cumings}}, \bibinfo {author} {\bibfnamefont {S.~J.}\ \bibnamefont {Gamble}},
  \bibinfo {author} {\bibfnamefont {R.~d.}\ \bibnamefont {Picciotto}}, \bibinfo
  {author} {\bibfnamefont {H.}~\bibnamefont {Kataura}}, \ and\ \bibinfo
  {author} {\bibfnamefont {D.}~\bibnamefont {Goldhaber-Gordon}},\ }\href
  {\doibase 10.1103/PhysRevB.76.245311} {\bibfield  {journal} {\bibinfo
  {journal} {Phys. Rev. B}\ }\textbf {\bibinfo {volume} {76}},\ \bibinfo
  {pages} {245311} (\bibinfo {year} {2007})}\BibitemShut {NoStop}%
\bibitem [{\citenamefont {Patton}\ \emph {et~al.}(2007)\citenamefont {Patton},
  \citenamefont {Kettemann}, \citenamefont {Zhuravlev},\ and\ \citenamefont
  {Lichtenstein}}]{Patton2007}%
  \BibitemOpen
  \bibfield  {author} {\bibinfo {author} {\bibfnamefont {K.~R.}\ \bibnamefont
  {Patton}}, \bibinfo {author} {\bibfnamefont {S.}~\bibnamefont {Kettemann}},
  \bibinfo {author} {\bibfnamefont {A.}~\bibnamefont {Zhuravlev}}, \ and\
  \bibinfo {author} {\bibfnamefont {A.}~\bibnamefont {Lichtenstein}},\ }\href
  {\doibase 10.1103/PhysRevB.76.100408} {\bibfield  {journal} {\bibinfo
  {journal} {Phys. Rev. B}\ }\textbf {\bibinfo {volume} {76}},\ \bibinfo
  {pages} {100408} (\bibinfo {year} {2007})}\BibitemShut {NoStop}%
\bibitem [{\citenamefont {Seridonio}\ \emph {et~al.}(2009)\citenamefont
  {Seridonio}, \citenamefont {Souza},\ and\ \citenamefont
  {Shelykh}}]{Seridonio2009}%
  \BibitemOpen
  \bibfield  {author} {\bibinfo {author} {\bibfnamefont {A.~C.}\ \bibnamefont
  {Seridonio}}, \bibinfo {author} {\bibfnamefont {F.~M.}\ \bibnamefont
  {Souza}}, \ and\ \bibinfo {author} {\bibfnamefont {I.~A.}\ \bibnamefont
  {Shelykh}},\ }\href@noop {} {\bibfield  {journal} {\bibinfo  {journal} {J.
  Phys.: Condens. Matter}\ }\textbf {\bibinfo {volume} {21}},\ \bibinfo {pages}
  {095003} (\bibinfo {year} {2009})}\BibitemShut {NoStop}%
\bibitem [{\citenamefont {Fu}\ \emph {et~al.}(2012)\citenamefont {Fu},
  \citenamefont {Xue},\ and\ \citenamefont {Wiesendanger}}]{Fu2012}%
  \BibitemOpen
  \bibfield  {author} {\bibinfo {author} {\bibfnamefont {Y.-S.}\ \bibnamefont
  {Fu}}, \bibinfo {author} {\bibfnamefont {Q.-K.}\ \bibnamefont {Xue}}, \ and\
  \bibinfo {author} {\bibfnamefont {R.}~\bibnamefont {Wiesendanger}},\ }\href
  {\doibase 10.1103/PhysRevLett.108.087203} {\bibfield  {journal} {\bibinfo
  {journal} {Phys. Rev. Lett.}\ }\textbf {\bibinfo {volume} {108}},\ \bibinfo
  {pages} {087203} (\bibinfo {year} {2012})}\BibitemShut {NoStop}%
\bibitem [{\citenamefont {Appelbaum}(1967)}]{Appelbaum1967}%
  \BibitemOpen
  \bibfield  {author} {\bibinfo {author} {\bibfnamefont {J.~A.}\ \bibnamefont
  {Appelbaum}},\ }\href {\doibase 10.1103/PhysRev.154.633} {\bibfield
  {journal} {\bibinfo  {journal} {Phys. Rev.}\ }\textbf {\bibinfo {volume}
  {154}},\ \bibinfo {pages} {633} (\bibinfo {year} {1967})}\BibitemShut
  {NoStop}%
\bibitem [{\citenamefont {Zhang}\ \emph {et~al.}(2013)\citenamefont {Zhang},
  \citenamefont {Kahle}, \citenamefont {Herden}, \citenamefont {Stroh},
  \citenamefont {Mayor}, \citenamefont {Schlickum}, \citenamefont {Ternes},
  \citenamefont {Wahl},\ and\ \citenamefont {Kern}}]{Zhang2013}%
  \BibitemOpen
  \bibfield  {author} {\bibinfo {author} {\bibfnamefont {Y.-h.}\ \bibnamefont
  {Zhang}}, \bibinfo {author} {\bibfnamefont {S.}~\bibnamefont {Kahle}},
  \bibinfo {author} {\bibfnamefont {T.}~\bibnamefont {Herden}}, \bibinfo
  {author} {\bibfnamefont {C.}~\bibnamefont {Stroh}}, \bibinfo {author}
  {\bibfnamefont {M.}~\bibnamefont {Mayor}}, \bibinfo {author} {\bibfnamefont
  {U.}~\bibnamefont {Schlickum}}, \bibinfo {author} {\bibfnamefont
  {M.}~\bibnamefont {Ternes}}, \bibinfo {author} {\bibfnamefont
  {P.}~\bibnamefont {Wahl}}, \ and\ \bibinfo {author} {\bibfnamefont
  {K.}~\bibnamefont {Kern}},\ }\href {\doibase 10.1038/ncomms3110} {\bibfield
  {journal} {\bibinfo  {journal} {Nature Commun.}\ }\textbf {\bibinfo {volume}
  {4}},\ \bibinfo {pages} {2110} (\bibinfo {year} {2013})}\BibitemShut
  {NoStop}%
\bibitem [{\citenamefont {Frota}(1992)}]{Frota.PhysRevB.45.1096}%
  \BibitemOpen
  \bibfield  {author} {\bibinfo {author} {\bibfnamefont {H.~O.}\ \bibnamefont
  {Frota}},\ }\href {\doibase 10.1103/PhysRevB.45.1096} {\bibfield  {journal}
  {\bibinfo  {journal} {Phys. Rev. B}\ }\textbf {\bibinfo {volume} {45}},\
  \bibinfo {pages} {1096} (\bibinfo {year} {1992})}\BibitemShut {NoStop}%
\bibitem [{\citenamefont {Madhavan}\ \emph {et~al.}(1998)\citenamefont
  {Madhavan}, \citenamefont {Chen}, \citenamefont {Jamneala}, \citenamefont
  {Crommie},\ and\ \citenamefont {Wingreen}}]{Madhavan1998}%
  \BibitemOpen
  \bibfield  {author} {\bibinfo {author} {\bibfnamefont {V.}~\bibnamefont
  {Madhavan}}, \bibinfo {author} {\bibfnamefont {W.}~\bibnamefont {Chen}},
  \bibinfo {author} {\bibfnamefont {T.}~\bibnamefont {Jamneala}}, \bibinfo
  {author} {\bibfnamefont {M.~F.}\ \bibnamefont {Crommie}}, \ and\ \bibinfo
  {author} {\bibfnamefont {N.~S.}\ \bibnamefont {Wingreen}},\ }\href@noop {}
  {\bibfield  {journal} {\bibinfo  {journal} {Sci}\ }\textbf {\bibinfo {volume}
  {280}},\ \bibinfo {pages} {567} (\bibinfo {year} {1998})}\BibitemShut
  {NoStop}%
\bibitem [{\citenamefont {Pr\"user}\ \emph {et~al.}(2011)\citenamefont
  {Pr\"user}, \citenamefont {Wenderoth}, \citenamefont {Dargel}, \citenamefont
  {Weismann}, \citenamefont {Peters}, \citenamefont {Pruschke},\ and\
  \citenamefont {Ulbrich}}]{Prueser2011}%
  \BibitemOpen
  \bibfield  {author} {\bibinfo {author} {\bibfnamefont {H.}~\bibnamefont
  {Pr\"user}}, \bibinfo {author} {\bibfnamefont {M.}~\bibnamefont {Wenderoth}},
  \bibinfo {author} {\bibfnamefont {P.~E.}\ \bibnamefont {Dargel}}, \bibinfo
  {author} {\bibfnamefont {A.}~\bibnamefont {Weismann}}, \bibinfo {author}
  {\bibfnamefont {R.}~\bibnamefont {Peters}}, \bibinfo {author} {\bibfnamefont
  {T.}~\bibnamefont {Pruschke}}, \ and\ \bibinfo {author} {\bibfnamefont
  {R.~G.}\ \bibnamefont {Ulbrich}},\ }\href {\doibase 10.1038/nphys1876}
  {\bibfield  {journal} {\bibinfo  {journal} {Nature Physics}\ }\textbf
  {\bibinfo {volume} {7}},\ \bibinfo {pages} {203} (\bibinfo {year}
  {2011})}\BibitemShut {NoStop}%
\bibitem [{\citenamefont {Heinrich}\ \emph {et~al.}(2004)\citenamefont
  {Heinrich}, \citenamefont {Gupta}, \citenamefont {Lutz},\ and\ \citenamefont
  {Eigler}}]{Heinrich2004}%
  \BibitemOpen
  \bibfield  {author} {\bibinfo {author} {\bibfnamefont {A.~J.}\ \bibnamefont
  {Heinrich}}, \bibinfo {author} {\bibfnamefont {J.}~\bibnamefont {Gupta}},
  \bibinfo {author} {\bibfnamefont {C.~P.}\ \bibnamefont {Lutz}}, \ and\
  \bibinfo {author} {\bibfnamefont {D.~M.}\ \bibnamefont {Eigler}},\ }\href
  {\doibase 10.1126/science.1101077} {\bibfield  {journal} {\bibinfo  {journal}
  {Science}\ }\textbf {\bibinfo {volume} {306}},\ \bibinfo {pages} {466 }
  (\bibinfo {year} {2004})}\BibitemShut {NoStop}%
\bibitem [{\citenamefont {Hirjibehedin}\ \emph {et~al.}(2006)\citenamefont
  {Hirjibehedin}, \citenamefont {Lutz},\ and\ \citenamefont
  {Heinrich}}]{Hirjibehedin2006}%
  \BibitemOpen
  \bibfield  {author} {\bibinfo {author} {\bibfnamefont {C.~F.}\ \bibnamefont
  {Hirjibehedin}}, \bibinfo {author} {\bibfnamefont {C.~P.}\ \bibnamefont
  {Lutz}}, \ and\ \bibinfo {author} {\bibfnamefont {A.~J.}\ \bibnamefont
  {Heinrich}},\ }\href {\doibase 10.1126/science.1125398} {\bibfield  {journal}
  {\bibinfo  {journal} {Science}\ }\textbf {\bibinfo {volume} {312}},\ \bibinfo
  {pages} {1021} (\bibinfo {year} {2006})}\BibitemShut {NoStop}%
\bibitem [{\citenamefont {Khajetoorians}\ \emph {et~al.}(2010)\citenamefont
  {Khajetoorians}, \citenamefont {Chilian}, \citenamefont {Wiebe},
  \citenamefont {Schuwalow}, \citenamefont {Lechermann},\ and\ \citenamefont
  {Wiesendanger}}]{Khajetoorians.N2010}%
  \BibitemOpen
  \bibfield  {author} {\bibinfo {author} {\bibfnamefont {A.~A.}\ \bibnamefont
  {Khajetoorians}}, \bibinfo {author} {\bibfnamefont {B.}~\bibnamefont
  {Chilian}}, \bibinfo {author} {\bibfnamefont {J.}~\bibnamefont {Wiebe}},
  \bibinfo {author} {\bibfnamefont {S.}~\bibnamefont {Schuwalow}}, \bibinfo
  {author} {\bibfnamefont {F.}~\bibnamefont {Lechermann}}, \ and\ \bibinfo
  {author} {\bibfnamefont {R.}~\bibnamefont {Wiesendanger}},\ }\href {\doibase
  10.1038/nature09519} {\bibfield  {journal} {\bibinfo  {journal} {Nature}\
  }\textbf {\bibinfo {volume} {467}},\ \bibinfo {pages} {1084} (\bibinfo {year}
  {2010})}\BibitemShut {NoStop}%
\bibitem [{\citenamefont {Khajetoorians}\ \emph {et~al.}(2011)\citenamefont
  {Khajetoorians}, \citenamefont {Lounis}, \citenamefont {Chilian},
  \citenamefont {Costa}, \citenamefont {Zhou}, \citenamefont {Mills},
  \citenamefont {Wiebe},\ and\ \citenamefont
  {Wiesendanger}}]{Khajetoorians.PRL2011}%
  \BibitemOpen
  \bibfield  {author} {\bibinfo {author} {\bibfnamefont {A.~A.}\ \bibnamefont
  {Khajetoorians}}, \bibinfo {author} {\bibfnamefont {S.}~\bibnamefont
  {Lounis}}, \bibinfo {author} {\bibfnamefont {B.}~\bibnamefont {Chilian}},
  \bibinfo {author} {\bibfnamefont {A.~T.}\ \bibnamefont {Costa}}, \bibinfo
  {author} {\bibfnamefont {L.}~\bibnamefont {Zhou}}, \bibinfo {author}
  {\bibfnamefont {D.~L.}\ \bibnamefont {Mills}}, \bibinfo {author}
  {\bibfnamefont {J.}~\bibnamefont {Wiebe}}, \ and\ \bibinfo {author}
  {\bibfnamefont {R.}~\bibnamefont {Wiesendanger}},\ }\href {\doibase
  10.1103/PhysRevLett.106.037205} {\bibfield  {journal} {\bibinfo  {journal}
  {Physical Review Letters}\ }\textbf {\bibinfo {volume} {106}},\ \bibinfo
  {pages} {037205} (\bibinfo {year} {2011})}\BibitemShut {NoStop}%
\bibitem [{\citenamefont {Kahle}\ \emph {et~al.}(2012)\citenamefont {Kahle},
  \citenamefont {Deng}, \citenamefont {Malinowski}, \citenamefont {Tonnoir},
  \citenamefont {Forment-Aliaga}, \citenamefont {Thontasen}, \citenamefont
  {Rinke}, \citenamefont {Le}, \citenamefont {Turkowski}, \citenamefont
  {Rahman}, \citenamefont {Rauschenbach}, \citenamefont {Ternes},\ and\
  \citenamefont {Kern}}]{Kahle2012}%
  \BibitemOpen
  \bibfield  {author} {\bibinfo {author} {\bibfnamefont {S.}~\bibnamefont
  {Kahle}}, \bibinfo {author} {\bibfnamefont {Z.}~\bibnamefont {Deng}},
  \bibinfo {author} {\bibfnamefont {N.}~\bibnamefont {Malinowski}}, \bibinfo
  {author} {\bibfnamefont {C.}~\bibnamefont {Tonnoir}}, \bibinfo {author}
  {\bibfnamefont {A.}~\bibnamefont {Forment-Aliaga}}, \bibinfo {author}
  {\bibfnamefont {N.}~\bibnamefont {Thontasen}}, \bibinfo {author}
  {\bibfnamefont {G.}~\bibnamefont {Rinke}}, \bibinfo {author} {\bibfnamefont
  {D.}~\bibnamefont {Le}}, \bibinfo {author} {\bibfnamefont {V.}~\bibnamefont
  {Turkowski}}, \bibinfo {author} {\bibfnamefont {T.~S.}\ \bibnamefont
  {Rahman}}, \bibinfo {author} {\bibfnamefont {S.}~\bibnamefont
  {Rauschenbach}}, \bibinfo {author} {\bibfnamefont {M.}~\bibnamefont
  {Ternes}}, \ and\ \bibinfo {author} {\bibfnamefont {K.}~\bibnamefont
  {Kern}},\ }\href@noop {} {\bibfield  {journal} {\bibinfo  {journal} {Nano
  Lett.}\ }\textbf {\bibinfo {volume} {12}},\ \bibinfo {pages} {518} (\bibinfo
  {year} {2012})}\BibitemShut {NoStop}%
\bibitem [{\citenamefont {Wiesendanger}(2009)}]{Wiesendanger2009}%
  \BibitemOpen
  \bibfield  {author} {\bibinfo {author} {\bibfnamefont {R.}~\bibnamefont
  {Wiesendanger}},\ }\href {\doibase 10.1103/RevModPhys.81.1495} {\bibfield
  {journal} {\bibinfo  {journal} {Rev. Mod. Phys.}\ }\textbf {\bibinfo {volume}
  {81}},\ \bibinfo {pages} {1495} (\bibinfo {year} {2009})}\BibitemShut
  {NoStop}%
\bibitem [{\citenamefont {Loth}\ \emph
  {et~al.}(2010{\natexlab{a}})\citenamefont {Loth}, \citenamefont {von
  Bergmann}, \citenamefont {Ternes}, \citenamefont {Otte}, \citenamefont
  {Lutz},\ and\ \citenamefont {Heinrich}}]{Loth.NP2010}%
  \BibitemOpen
  \bibfield  {author} {\bibinfo {author} {\bibfnamefont {S.}~\bibnamefont
  {Loth}}, \bibinfo {author} {\bibfnamefont {K.}~\bibnamefont {von Bergmann}},
  \bibinfo {author} {\bibfnamefont {M.}~\bibnamefont {Ternes}}, \bibinfo
  {author} {\bibfnamefont {A.~F.}\ \bibnamefont {Otte}}, \bibinfo {author}
  {\bibfnamefont {C.~P.}\ \bibnamefont {Lutz}}, \ and\ \bibinfo {author}
  {\bibfnamefont {A.~J.}\ \bibnamefont {Heinrich}},\ }\href {\doibase
  10.1038/nphys1616} {\bibfield  {journal} {\bibinfo  {journal} {Nature
  Physics}\ }\textbf {\bibinfo {volume} {6}},\ \bibinfo {pages} {340} (\bibinfo
  {year} {2010}{\natexlab{a}})}\BibitemShut {NoStop}%
\bibitem [{\citenamefont {Zitko}\ \emph {et~al.}(2009)\citenamefont {Zitko},
  \citenamefont {Peters},\ and\ \citenamefont {Pruschke}}]{Zitko2009}%
  \BibitemOpen
  \bibfield  {author} {\bibinfo {author} {\bibfnamefont {R.}~\bibnamefont
  {Zitko}}, \bibinfo {author} {\bibfnamefont {R.}~\bibnamefont {Peters}}, \
  and\ \bibinfo {author} {\bibfnamefont {T.}~\bibnamefont {Pruschke}},\
  }\href@noop {} {\bibfield  {journal} {\bibinfo  {journal} {New Journal of
  Physics}\ }\textbf {\bibinfo {volume} {11}},\ \bibinfo {pages} {053003}
  (\bibinfo {year} {2009})}\BibitemShut {NoStop}%
\bibitem [{\citenamefont {Loth}\ \emph
  {et~al.}(2010{\natexlab{b}})\citenamefont {Loth}, \citenamefont {Lutz},\ and\
  \citenamefont {Heinrich}}]{Loth.NJP.2010}%
  \BibitemOpen
  \bibfield  {author} {\bibinfo {author} {\bibfnamefont {S.}~\bibnamefont
  {Loth}}, \bibinfo {author} {\bibfnamefont {C.~P.}\ \bibnamefont {Lutz}}, \
  and\ \bibinfo {author} {\bibfnamefont {A.~J.}\ \bibnamefont {Heinrich}},\
  }\href@noop {} {\bibfield  {journal} {\bibinfo  {journal} {New J. Phys.}\
  }\textbf {\bibinfo {volume} {12}},\ \bibinfo {pages} {125021} (\bibinfo
  {year} {2010}{\natexlab{b}})}\BibitemShut {NoStop}%
\bibitem [{\citenamefont {Meservey}\ and\ \citenamefont
  {Tedrow}(1994)}]{Meservey1994}%
  \BibitemOpen
  \bibfield  {author} {\bibinfo {author} {\bibfnamefont {R.}~\bibnamefont
  {Meservey}}\ and\ \bibinfo {author} {\bibfnamefont {P.}~\bibnamefont
  {Tedrow}},\ }\href@noop {} {\bibfield  {journal} {\bibinfo  {journal} {Phys.
  Rep.}\ }\textbf {\bibinfo {volume} {238}},\ \bibinfo {pages} {173} (\bibinfo
  {year} {1994})}\BibitemShut {NoStop}%
\end{thebibliography}
\end{document}